\documentclass[11pt]{article}
\usepackage{moriond,epsfig}

\def\be{\begin{equation}}
\def\ee{\end{equation}}
\def\bea{\begin{eqnarray}}
\def\eea{\end{eqnarray}}

\newcommand{\Mpl}{M_\mathrm{pl}} 
\newcommand{\mpl}{m_\mathrm{pl}} 

\newcommand{\rr}{\mathrm}

\begin{document}
\title{INITIAL CONDITIONS IN HYBRID INFLATION:  EXPLORATION BY MCMC TECHNIQUE}

\author{SEBASTIEN CLESSE}

\address{Service de Physique Th\'eorique, Universit\'e Libre de Bruxelles, 
CP225, Boulevard du Triomphe, 1050 Brussels, Belgium \\
Center for Particle Physics and Phenomenology, Louvain University, 2 chemin du cyclotron, 1348 Louvain-la-Neuve, Belgium}

\maketitle\abstracts{
In hybrid inflation, initial field values leading to sufficiently long inflation were thought be fine-tuned in a narrow band along the inflationary valley.  A re-analysis of this problem has shown that there exists a non negligible proportion of successful initial conditions exterior to the valley, organized in a complex structure with fractal boundaries, and whose origin has been explained.  Their existence in a large part of the parameter space has been demonstrated using a bayesian Monte-Carlo-Markov-Chain (MCMC) method, and natural bounds on potential parameters have been established.  Moreover, these results are shown to be valid not only for the original hybrid model, but also for other hybrid realizations in various frameworks.  }

\section{Introduction and original hybrid model}

     Recent observations of the Cosmic Microwave Background (CMB) and its anisotropies have provided strong arguments in favour of a phase of accelerated expansion in the early universe.   If the simpler way to realize this inflation era is to assume the universe initially filled with an unique homogeneous scalar field slowly rolling along its potential, many other realizations have been proposed \footnote{see e.g.\cite{Mazumdar:2010sa} for a recent review.}.   In hybrid models, an inflaton field is coupled to an auxiliary field, and accelerated expansion usually occurs when inflaton slow-rolls along a nearly flat valley of the potential and ends abruptly due to a tachyonic instability.   The advantages of hybrid models are that the  energy scale of inflation can be low and  initial field values do not need to be larger than the Planck mass.  Moreover, the model can be embedded in some high energy frameworks like supersymmetry, supergravity and grand unified theories.   
A generic prediction of the original hybrid model \cite{Linde:1993cn,Copeland:1994vg} is a blue spectrum of scalar perturbations, and thus it is strongly disfavoured by WMAP constraints on the scalar spectral index \cite{Martin:2006rs}.  It remains nevertheless a good toy model for other hybrid realizations in various frameworks, whose dynamics is similar.

The potential for the original hybrid model reads \cite{Linde:1993cn,Copeland:1994vg}
\begin{equation}
V(\phi,\psi) = \Lambda^4 \left[ \left(1- \frac {\psi^2}{M^2} \right)^2 + \frac{\phi^2}{\mu^2} + 2 \frac{\phi^2 \psi^2}{\phi_{\rr c}^2 M^2}   \right]
\end{equation}
in which $\phi$ is the inflaton, $\psi$ is the auxiliary Higgs-type field, and $M, \mu, \phi_{\rr c} $ are three mass parameters.  Inflation occurs in the false-vacuum along the nearly flat valley $\psi = 0$.  A tachyonic instability appears when inflaton reaches $\phi = \phi_{\rr c} $.  From this point, the trajectory falls through one of the global minima of the potential $\phi=0, \psi = \pm M$   whereas tachyonic preheating occurs \cite{Felder:2000hj}.   

It is a natural question to ask how the fields have to be tuned initially along the inflationary valley in order to generate more than around 60 e-folds of inflation inside the valley. If trajectories are initially displaced slightly in the transverse direction, they perform damped oscillations along the valley before the slow-roll regime engages the realization of a large number of e-folds.  On the contrary, if the initial value of the auxiliary field $\psi_{\rr i}$ is too large,  the damping can not proceed sufficiently quickly and slow-roll regime do not begin before the critical point of instability is reached. Tetradis \cite{Tetradis:1997kp} and afterward Mendes and Liddle \cite{Mendes:2000sq}  determined that the successful region in the initial field space was a very narrow band along the valley such that the initial value of the auxiliary field had to be fine-tuned compared to initial inflaton values.   However, some inconsistencies can be pointed out and have lead to the reanalysis of the problem in \cite{Clesse:2009ur,Clesse:2008pf,Clesse:2009zd} .   Both \cite{Tetradis:1997kp} and \cite{Mendes:2000sq}   observed some unexplained successful initial conditions outside the valley, who seem organised in some structures in \cite{Tetradis:1997kp}, but apparently isolated and very subdominant in the higher-resolution analysis of \cite{Mendes:2000sq} .  These previous studies were also restricted to some similar sets of potential parameters.  A full quantitative study of the problem of initial field values in the whole parameter space was still lacking.    

By integrating numerically the classical 2-field dynamics, the behaviour of such successful trajectories starting outside the valley can be understood.   The relative area that these points can occupy can be evaluated for some sets of parameters, and their dominance in the whole space of potential parameters, initial field-values and initial velocities can be probed using a statistical Monte-Carlo-Markov-Chain method.  Our results indicate that the model do not suffer of fine-tuning any more because successful trajectories start  more probably outside the valley.  It is also shown that initial velocities do not affect the probability to generate sufficient inflation.  Finally, some bounds on potential parameters can be established with the only requirement of sufficiently long period of inflation and the absence of some fine-tuning of initial conditions.   The important questions of the robustness of these observations for other hybrid realizations, as well as the effects of quantum fluctuations on such trajectories can be pointed out and will be addressed briefly in the conclusion.  

\section{Space of initial conditions}

\subsection{Grids of initial field values for fixed parameters}

\begin{figure}[]  
\hspace{40mm}
 \includegraphics[width=90mm]{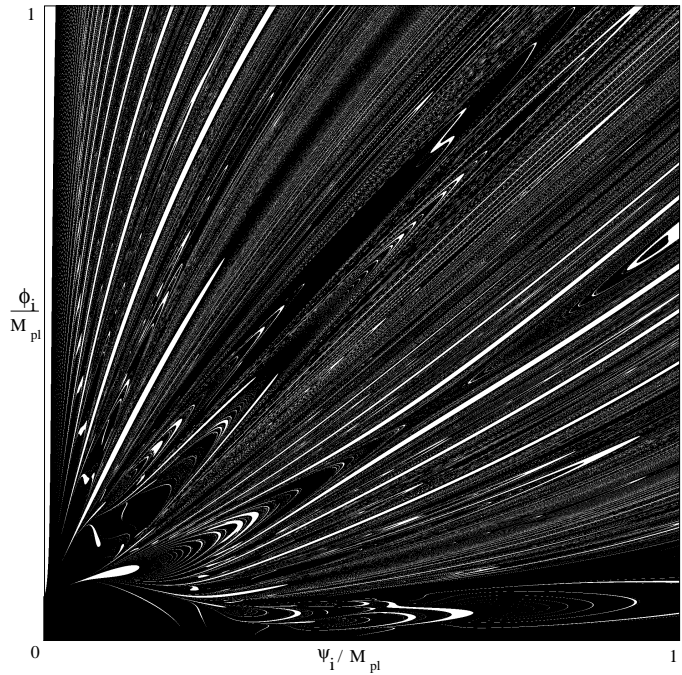} 


 \caption{grid (2040x2040 points) of initial field values producing more/less than $60$ e-folds of
    inflation (white/black regions). The potential parameters are set
    to $M=0.03 \  \mpl$, $\mu = 636 \ \mpl$, $\phi_{\rr c} = \sqrt 2 \times 10^{-2}  \ \mpl$.}
 
  \label{fig:grid}
\end{figure}

High-resolution grids of initial field values have been plotted by integrating numerically the classical 2-field dynamics.  In a flat FLRW universe, dynamics is governed by Friedmann-Lema\^{\i}tre equations 
\begin{equation} \label{eq:FLtc12field}
H^2 = \frac {8\pi }{3 \mpl^2}  \left[ \frac 1 2 \left(\dot
\phi^2 + \dot \psi^2 \right)  + V(\phi,\psi) \right], \hspace{10mm}
\frac{\ddot a }{a} = \frac {8\pi}{3 \mpl^2} \left[ - \dot \phi^2
- \dot \psi^2 + V(\phi,\psi ) \right]~,
\end{equation}
as well as Klein-Gordon equations in an expanding universe
\begin{equation} \label{eq:KGtc2field}
\ddot \phi + 3 H \dot \phi + \frac {\partial
V(\phi,\psi)}{\partial \phi} = 0~, \hspace{10mm}
\ddot \psi + 3 H \dot \psi + \frac {\partial 
V(\phi,\psi)}{\partial \psi} = 0~,
\end{equation}
where a dot denotes derivative with respect to cosmic time, $\mpl$ is the Planck mass \footnote{$\Mpl$ is used for the reduced Planck mass $\mpl / \sqrt{8 \pi}$ }$ H \equiv \dot a / a $ and $a$ is the scale factor. In a first step, initial velocities are assumed to be vanishing.  White regions in the grid represented in figure \ref{fig:grid} correspond to initial conditions leading to more than $60$ e-folds of inflation.  As expected, a band of fine-tuned initial field values is found along the valley, as well as successful points outside the valley.  But instead of being isolated, as it was suggested in \cite{Mendes:2000sq} , they form a complex structure of thin lines and crescents, and with successive zooming  it is observed that the structure form a connected set.   Moreover, the fractal properties of these successful initial conditions have been studied \cite{Clesse:2009ur}.  The box-counting dimension of the boundaries has been determined numerically and found to be non-integer.   Nevertheless, the structure itself exhibits a non-fractal finite area.  These properties are similar to the well known Mandelbrot set whose area is finite with infinite fractal boundaries.  

Finally, the relative area covered by successful points have been quantified.  These points are found to cover up to 20\% of the space of initial field values, depending on potential parameter sets \cite{Clesse:2008pf}.  Therefore, the amount of fine-tuning necessary to have successful inflation is strongly reduced for these particular sets of parameters.     

\subsection{Successful trajectories initially outside the valley}

\begin{figure}[h!]
\hspace{35mm}  
   \includegraphics[width=10cm]{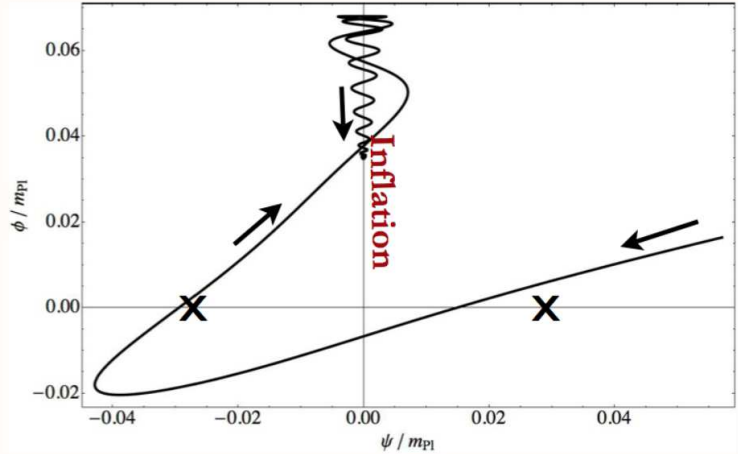}

  \caption{Typical behaviour of successful trajectory starting outside the inflationary valley, first falling towards the bottom of the potential, then oscillating and climbing the inflationary valley ($\psi =0$ direction) instead of being trapped by one of the global minima of the potential (represented by crosses).  Inflation occurs when it slow-rolls back along the valley.}
  \label{fig:trajectory}
\end{figure}

The typical behaviour of successful trajectories starting outside the inflationary valley (see fig.\ref{fig:trajectory}) has been studied more precisely.   Three successive phases have been identified.  First, a fast-rolling phase pushing the trajectories through the bottom of the potential, without generating a significant number of e-folds.   Then, after some rebonds on the sides of the potential, these trajectories become oriented along the valley, and climb it.   Arriving at an extremum point with a quasi-vanishing velocity \footnote{This extremum point is located inside the fine-tuned successful band, thus there exists a correspondence between each initial value inside and exterior to the valley.}, they start to slow roll back along the valley while producing a large number of e-folds.  

\subsection{Statistical MCMC analysis}

Grids of initial conditions demonstrates that a severe fine-tuning of initial field values is not required in hybrid inflation for particular sets of potential parameters.   However, the full initial fields and potential parameter space including initial velocities is 7-th dimensional \footnote{Let notice that $\Lambda$ has not been considered since it only normalizes the potential without influencing the dynamics.} and thus is hardly explored with such grids.   For this reason, a statistical bayesian MCMC analysis of this space has been performed.    Priors are chosen to be flat for initial field values, initial velocities, and for the logarithm of the potential parameters, in order to avoid fine-tuning of the initial fields and to not favour any precise scale for the potential parameters.    Only the sub-Planckian regime has been considered, but it was shown in \cite{Clesse:2008pf}  that for super-Planckian trajectories, the realization of at least 60 e-folds of inflation becomes generic.  More details about the MCMC method can be found in \cite{Clesse:2009ur}.  Only the main results will be discussed below.  

\begin{figure}[h!]
\hspace{40mm}  
   \includegraphics[width=7cm]{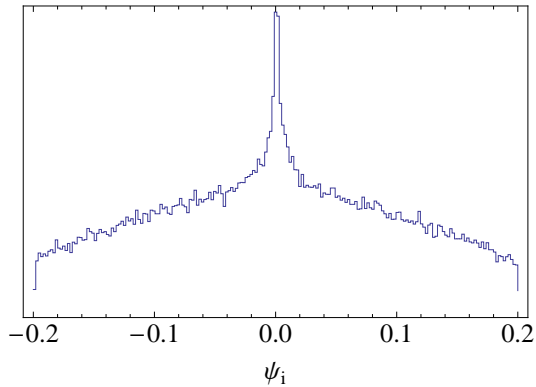}

  \caption{Posterior probability density distribution for $\psi_{\rr i}$ (in $\mpl$ units) with the requirement of at least 60 e-folds generated, marginalized over initial values of $\phi$, potential parameters and initial field velocities.}
  \label{fig:probapsi}
\end{figure}

The first result concerns the posterior probability density distribution of $\psi_{\rr i} $ marginalized over the rest of the space (see figure \ref{fig:probapsi}).   It exhibits a maximum at $\psi_{\rr i} = 0$ corresponding to successful trajectories starting inside the valley.  However, it is important to remark that the probability distribution is widely spread over large values of  $\psi_{\rr i} $, so that the probability to have a successful trajectory starting exterior to the valley is dominant over the probability to start along it.  It can therefore be concluded that in the original hybrid model, there is  no need of fine-tuned initial conditions along the valley to generate sufficient inflation. 
Concerning initial velocities, posterior probability distributions are flat, and they do not play a role in the game of finding trajectories leading to many e-folds of inflation.   
Finally, from the only requirement of having more than 60 e-folds of inflation, natural bounds on the potential parameters $\phi_{\rr c}$ and $\mu$  can be established, 
\begin{equation} 
\phi_{\rr c} < 0.004 \  \mpl  \ \rr{95 \% C.L.}, \hspace{10mm} \mu > 1.7 \ \mpl \  \rr{95 \% C.L.}.
\end{equation}
From the first bound, inflaton value at instability point is shown to be smaller than the reduced planck mass, and the last 60 e-folds are sure to be generated in the sub-planckian regime inside the valley. The second bound is linked to the non-existence of a small field phase of inflation along the valley for small values of $\mu$.  This regime was put in evidence in \cite{Clesse:2008pf}.

\section{Conclusion and discussion}

By integrating the classical 2-field dynamics, it has been shown that the original hybrid model does not require fine-tuned initial field values in order to generate a sufficient number of e-folds to solve the standard cosmological problems.   Actually, trajectories do not need to be initially tuned inside a narrow band along the inflationary valley.  They can start exterior to the valley, and after a fast-roll phase and some rebonds on the sides of the potential, they join it and enter in the slow-roll regime.   A bayesian MCMC method has been used to show that this observation is valid, not only for some specific potential parameter sets, but in a large part of the 7-D space of initial field values, initial velocities and potential parameters.    Initial velocities do not influence the probability to generate sufficient inflation, and bounds on potential parameters have been deduced from the only requirement of a sufficiently long period of inflation.  

If the analysis of the successful initial conditions has been conducted essentially for original hybrid model, the results remain generic for other hybrid realizations in various framework \cite{Clesse:2008pf,Clesse:2009ur} .  A large part of the space of initial field values (up to 80\%) has been found to be successful for supersymmetric smooth \cite{Lazarides:1995vr} and shifted \cite{Jeannerot:2000sv,Jeannerot:2002wt} inflation, both in their SUSY and SUGRA versions, as well as for radion assisted gauge inflation \cite{Fairbairn:2003yx}.  The MCMC analysis have been conducted for F-term hybrid inflation \cite{Dvali:1994ms} in supergravity, with similar results.  In particular, a bound on the only parameter $M$ of the model has been established,
\begin{equation}
\log (M / \Mpl) < -1.33 \ \rr{95 \% \ C.L.} ,
\end{equation}
comparable to the upper bound obtained from cosmic strings formation in \cite{Jeannerot:2005mc}.  

One could argue that there is a remaining problem in the sense that  successful initial conditions are subdominant (they do not exceed 25\% of the space of initial field values for original hybrid model) compared to unsuccessful trajectories.  However, it is without noticing that successful points outside the valley form a complex fractal structure covering roughly all parts of the space of initial conditions.   In the context of self-reproducing universe, due to quantum fluctuations along trajectories, each initial patch of the universe could contain at least one region for which inflation will undergo while in other regions inflation will not occur.  At the end, the volume occupied by the regions in which inflation occurred is largely dominant.  This effect should be studied more in details in a future work.

\section*{Acknowledgments}

This work has been done in collaboration with J. Rocher and C. Ringeval.  It is a pleasure to thank T. Carletti, A. Fuzfa, A. Lemaitre and M. Tytgat for fruitful discussions and comments.  S.C. is supported by the Belgian Fund for research F.R.I.A. 

\bibliography{biblio}

\end{document}